# Design and analysis of continuous hybrid differentiator

Xinhua Wang and Hai Lin

Department of Electrical & Computer Engineering, National University of Singapore, 4 Engineering Drive 3, Singapore 117576

E-mail: wangxinhua04@gmail.com

**Abstract:** In this paper, a continuous hybrid differentiator is presented based on a strong Lyapunov function. The differentiator design can not only reduce sufficiently chattering phenomenon of derivative estimation by introducing a perturbation parameter, but also the dynamical performances are improved by adding linear correction terms to the nonlinear ones. Moreover, strong robustness ability is obtained by integrating sliding mode items and the linear filter. Frequency analysis is applied to compare the hybrid continuous differentiator with sliding mode differentiator. The merits of the continuous hybrid differentiator include the excellent dynamical performances, restraining noises sufficiently, and avoiding the chattering phenomenon.

**Keywords:** continuous hybrid, differentiator, frequency analysis.

## 1. Introduction

Differentiation of signals is a well-known problem [1-6], which has attracted much attention in recent years. Obtaining the velocities of tracked targets is crucial for several kinds of systems with correct and timely performances, such as the missile-interception systems [7] and underwater vehicle systems [8], in which disturbances must be restrained. The simple realization and robustness of differentiators should be taken into consideration.

The high-gain differentiators [4, 19, 22] provide for an exact derivative when their gains tend to infinity. Unfortunately, their sensitivity to small high-frequency noise also infinitely grows. With any finite gain values such a differentiator has also a finite bandwidth. Thus, being not exact, it is, at the same time, insensitive with respect to high-frequency noise. Such insensitivity may be considered both as advantage or disadvantage depending on the circumstances. Moreover, high gain results in peaking phenomenon.

In [5, 6], a differentiator via second-order (or high-order) sliding modes algorithm has been proposed, and its frequency characteristics is given in [26]. The information one needs to know on the signal is an upper bound for Lipschitz constant of the derivative of the signal. Although second-order sliding mode is introduced and there exists no chattering phenomenon in signal tracking. However, for the robust exact differentiator, in the second dynamical equation, a switching function exists. The output of derivative estimation is continuous but not smooth. Therefore, chattering phenomenon still exists in derivative estimation.

In [28, 29], a switching function is designed to switches between the robust exact differentiator and a standard lead filter so as to render the error system uniformly globally exponentially practically stable without considering the existence of noises. Thus a global robust exact differentiator (GRED) is integrated. The switching function is decided by the outputs error of the two differentiators and some given parameters. When noises exist in signal, the exact derivative estimation cannot be obtained by these two differentiators independently or by GRED. It is difficult to select these parameters of the switching function when noises exist. Moreover, though no chattering phenomenon happens for signal tracking, it still exists in derivative estimation because of discontinuous function in the second equation of robust exact differentiator.

In [18], we presented a finite-time-convergent differentiator based on finite-time stability [15-17] and singular perturbation technique [9-14]. However, the differentiators in [18] are complicated. In [20], we designed a hybrid differentiator with high speed convergence, and it succeeds in applications to velocities estimation for low-speed regions only based on position measurements [21] and to velocities estimation for a quadrotor aircraft [27], in which only the convergence of signal tracking was described for this differentiator, but the convergence of derivative estimation was not given, and the regulation of parameters has no rules. Moreover, the proposed differentiator requires the boundedness of the first-order and second-order derivatives of the input signal.

From the previous analysis of linear differentiators [4, 19, 22] and sliding mode differentiators [5, 6], the linear differentiator and the nonlinear differentiator, two notable differences can be observed. First, the two algorithms have quite different converging properties: the linear system converges exponentially, whereas the trajectories of the sliding mode algorithm converge in finite time. This is due to the lack of local Lipschitzness of the SOSM algorithm at the origin, that is, its behavior around the zero state is very strong compared to the linear case. On the other side, the linear correction terms are stronger than the ones of the sliding mode algorithm far from the origin. These differences cause another striking difference between both algorithms: the kind of perturbations that each one is able to tolerate. The main difference is that the linear system can deal with perturbations that are stronger very far from the origin and weaker near the origin than the ones that are endured by the sliding mode algorithm. So, for example, the sliding mode algorithm is not able to endure (globally) a bounded perturbation with linear growth in time, but the linear algorithm can deal with it easily. However, the linear algorithm is not able to support a strong perturbation near the origin, what is one of the main advantages of the sliding mode.

In this paper, in order to integrate the merits of linear differentiator and robust exact differentiator, and to restrain their shortcomings respectively, a continuous hybrid differentiator algorithm is presented based on strong Lyapunov function. The proposed differentiator is an integration of a nonlinear term (comprising of continuous power function) and a linear correction term. The overall design is an extension of the second order differentiators proposed by Arie Levant [5, 6] in that a perturbation parameter α is introduced which takes values (0,1) and an additional linear correction term is appended, so that it inherits the best properties of both. The switching function in the second equation of robust exact differentiator is substituted by a continuous power function with a perturbation parameter. Therefore, chattering phenomenon can be avoided in the output of derivative estimation. Strong robustness ability by Lyapunov function is obtained by integrating sliding mode items and the linear filter. The linear part can restrain high-frequency noises by giving a suitable nature frequency, and small bounded noises can be restrained by the sliding mode items, at the same time, the sliding mode items can compensate the delay brought by the linear filter. Moreover, the proposed continuous hybrid differentiator only requires the boundness of the second-order derivative of the input signal.

A frequency response method, describing function method [13, 23], can be used to approximately analyze and predict nonlinear behaviors of differentiators. Frequency analysis is applied to compare hybrid continuous differentiator with sliding mode differentiator. Moreover, the advantage of the use of Lyapunov functions is that it is possible to obtain explicit relations for the design parameters.

## 2. Preliminaries

First of all, the concepts related to finite-time control are given.

**Definition 1 [16]:** Consider a time-invariant system in the form of

$$\dot{x} = f(x), \quad f(0) = 0, \quad x \in R^n \tag{1}$$

where $f : \hat{U}_0 \to R^n$ is continuous on an open neighborhood $\hat{U}_0$ of the origin. The equilibrium $x = 0$ of the system is (locally) finite-time stable if (i) it is asymptotically stable, in $\hat{U}$, an open neighborhood of the origin, with $\hat{U} \subseteq \hat{U}_0$; (ii) it is finite-time convergent in $\hat{U}$, that is, for any initial condition $x_0 \in \hat{U} \setminus \{0\}$, there is a settling time $T > 0$ such that every solution $x(t, x_0)$ of system (1) is defined with $x(t, x_0) \in \hat{U} \setminus \{0\}$ for $t \in [0, T]$ and satisfies

$$\lim_{t \to T} x(t, x_0) = 0 \tag{2}$$

and $x(t, x_0) = 0$, if $t \geq T$. Moreover, if $\hat{U} = R^n$, the origin $x = 0$ is globally finite-time stable.

**Definition 2 [31, 32, 33]:** A family of dilations $\delta_\rho^r$ is a mapping that assigns to every real $\rho > 0$ a diffeomorphism

$$\delta_\rho^r(x_1,\cdots,x_n)=\left(\rho^{r_1}x_1,\cdots,\rho^{r_n}x_n\right) \tag{3}$$

where $x_1,\cdots x_n$ are suitable coordinates on $R^n$ and $r=(r_1,\cdots,r_n)$ with the dilation coefficients $r_1\cdots,r_n$ positive real numbers.

A vector field $f(x)=(f_1(x),\cdots,f_n(x))^T$ is homogeneous of degree $k\in R$ with respect to the family of dilations $\delta_\rho^r$ if

$$f_i\left(\rho^{r_1}x_1,\cdots,\rho^{r_n}x_n\right)=\rho^{r_i+k}f_i(x), i=1,\cdots,n,\ \rho>0 \tag{4}$$

System (1) is called homogeneous if its vector field $f$ is homogeneous.

The following lemma was presented in some references like [15, 17, 24, 25].

**Lemma1** [15, 24, 25]**:** Suppose that system (1) is homogeneous of degree $k<0$ with respect to the family of dilations $\delta_\rho^r$, $f(x)$ is continuous and $x=0$ is its asymptotically stable equilibrium. Then equilibrium of system (1) is globally finite-time stable.

**Lemma 2 [17].** Suppose the origin is a finite-time-stable equilibrium [17, Theorem 4.3] of (1), and the settling-time function $T_f$ is continuous at zero, where $f()$ is continuous. Let $N$ be defined as in Definition 1 and let $\theta\in(0,1)$. Then there exists a continuous scalar function $V$ such that 1) $V$ is positive definite and 2) $dV/dt$ is real valued and continuous on $N$ and there exists $c>0$ such that

$$\dot{V}+cV^\theta\leq 0\ ,\qquad \text{for any}\ \ \theta\in(0,1) \tag{5}$$

**Assumption 1.** For (1), there exist $\rho\in(0,1]$ and a nonnegative constant $\bar{a}$ such that

$$\|f(z_1)-f(\bar{z}_1)\|\leq \bar{a}\|z_i-\bar{z}_i\|^\rho \tag{6}$$

where $z,\bar{z}\in\Re^n$.

**Remark 1.** There are a number of nonlinear functions actually satisfying assumption 2. For example, one such function is $x^\rho$ since $|x^\rho-\bar{x}^\rho|\leq 2^{1-\rho}|x-\bar{x}|^\rho, \rho\in(0,1]$. Moreover, there are smooth functions also satisfying this property. In fact, it is easy to verify that $|\sin x-\sin\bar{x}|\leq 2|x-\bar{x}|^\rho$ for any $\rho\in(0,1]$.

## 3. Problem statement

## A. Robust exact differentiator [5, 6]

In [5, 6], a finite-time convergent differentiator based on second-order sliding is presented as follow:

$$\dot{x}_{11} = x_{12} - \lambda_2 |x_{11} - v(t)|^{\frac{1}{2}} \operatorname{sgn}(x_{11} - v(t)) \tag{7}$$
$$\dot{x}_{12} = -\lambda_1 \operatorname{sgn}(x_{11} - v(t))$$

where the second-order differentiable signal $v(t)$ is bounded, and $|\ddot{v}(t)| \leq L_2$, $L_2$ is a positive constant, $\lambda_1, \lambda_2 > 0$, $\lambda_1 > L_2$. For differentiator (7), from Lemma 1, there exists a time $t_s > 0$ such that

$$x_{11} = v(t), x_{12} = \dot{v}(t) \tag{8}$$

for $t \geq t_s$.

Moreover, let $v(t)$ is the input signal with noise, $v_0(t)$ is the desired signal, and it is satisfied with $|v(t) - v_0(t)| \leq \varepsilon$. Therefore, for some positive constants $\mu_i$, $i = 0, 1$, the following inequalities are established [6]:

$$|e_{1,i+1}| = |x_{1,i+1} - v_0^i(t)| \leq \mu_i \varepsilon^{\frac{2-i}{2}}, \quad i = 0, 1 \tag{9}$$

In robust exact differentiator (7), though the output $x_{11}$ is smooth, the chattering phenomenon exists in the output $x_2$ because the discontinuous sgn( ) exists in the second differential equation of differentiator (7). In the motor velocity feedback systems, the chattering in $x_{12}$ can make motors trembling. Therefore, chattering phenomenon must be restrained sufficiently in the output $x_{12}$ for a differentiator. In the following, we analyse differentiator (7) from its frequency characteristics.

## A.1 Frequency characteristics of robust exact differentiator (7)

Let $e_{11} = x_{11} - v(t) = A \sin \omega t$, we have

$$\frac{2}{\pi} \int_0^\pi |A \sin \omega \tau|^{0.5} \operatorname{sgn}(A \sin \omega \tau) \sin \omega \tau d\omega t = A^{0.5} \frac{2}{\pi} \int_0^\pi |\sin \omega \tau|^{1.5} d\omega t \tag{10}$$

We can get $\frac{2}{\pi} \int_0^\pi \sin \omega \tau d\omega t = \frac{4}{\pi}$ and $\frac{2}{\pi} \int_0^\pi (\sin \omega \tau)^2 d\omega t = 1$.

In (10), let $\Omega = \frac{2}{\pi} \int_0^\pi |\sin \omega \tau|^{1.5} d\omega t$, therefore, we have $1 < \Omega < 4/\pi$. The describing function of nonlinear function $|\ |^{0.5} \operatorname{sgn}(\ )$ is $N(A) = \Omega / A^{0.5}$, and the linearization system of differentiator (7) is

$$\dot{x}_{11} = x_{12} - \lambda_2 \frac{\Omega}{A^{0.5}}(x_{11} - v(t))$$
$$\dot{x}_{12} = -\lambda_1 \frac{4}{\pi A}(x_{11} - v(t)) \tag{11}$$

The nature frequency of system (11) is

$$\omega_n = \frac{2\sqrt{\lambda_1}}{\sqrt{\pi} A^{0.5}} \tag{12}$$

and its damping coefficient is

$$\varsigma = \frac{\lambda_2 \Omega \sqrt{\pi}}{4\sqrt{\lambda_1}} \tag{13}$$

From (11), (12) and (13), we can find that when the tracking error is large, i.e., far from the origin, the gains in (11) are small. On the other hand, when the tracking error is small, i.e., around the origin, the gains are large. Therefore, the behavior of robust exact differentiator is weak far from the origin, and contrarily, the behavior around the origin is strong.

*A.2 Analysis of chattering phenomenon for robust exact differentiator (7)*

In the following, we give two examples to explain the chattering phenomenon in sliding control systems.

*Example 1:* First-order systems

1) First-order sliding mode system:

$$\dot{x} = u_1$$
$$u_1 = -2\,\mathrm{sgn}(x) \tag{14}$$

2) First-order linear system:

$$\dot{y} = u_2$$
$$u_2 = -2y \tag{15}$$

3) First-order continuous nonlinear system (with a power function):

$$\dot{z} = u_3$$
$$u_3 = -2|z|^{0.5}\,\mathrm{sgn}(z) \tag{16}$$

The simulation curves of the three systems are shown in Figures 1-3.

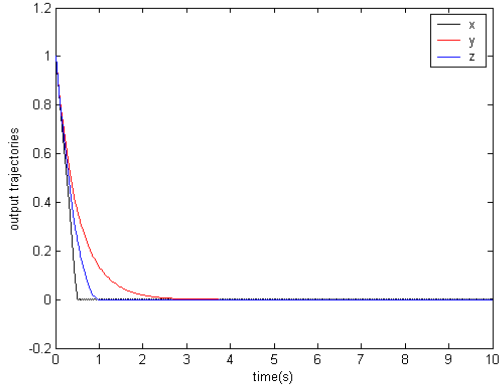 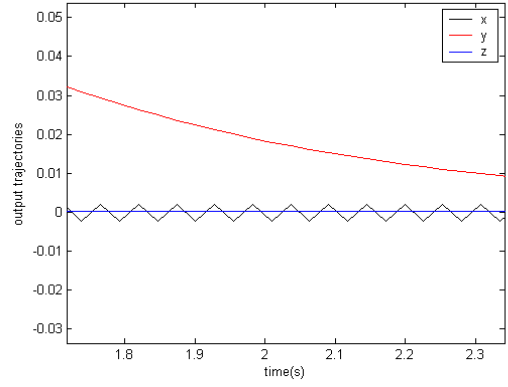

Fig. 1 $x$, $y$ and $z$          Fig. 2 The magnified figure of Fig. 1

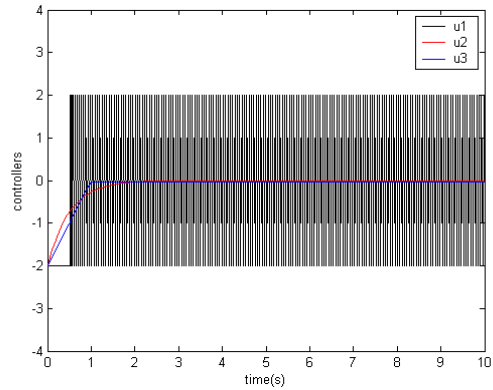

Fig. 3 $u_1$, $u_2$ and $u_3$

From Fig. 2, we can find that chattering phenomenon happen in system (14). And there is no chattering phenomenon in system (16).

*Example 2:* Second-order systems:

1) Second-order sliding mode system:

$$\begin{aligned} \dot{x}_1 &= x_2 - k_1 |x_{11}|^{\frac{1}{2}} \operatorname{sgn}(x_1) \\ \dot{x}_2 &= -k_2 \operatorname{sgn}(x_1) \end{aligned} \tag{17}$$

2) Second-order linear system:

$$\begin{aligned} \dot{y}_1 &= y_2 - k_1 y_1 \\ \dot{y}_2 &= -k_2 y_1 \end{aligned} \tag{18}$$

3) Second-order continuous nonlinear system:

$$\dot{z}_1 = z_2 - k_1 |z_1|^{\frac{\alpha+1}{2}} \operatorname{sgn}(z_1), \quad \alpha \in (0,1), \quad k_1, k_2 > 0 \tag{19}$$
$$\dot{z}_2 = -k_2 |z_1|^{\alpha} \operatorname{sgn}(z_1)$$

For second-order continuous nonlinear system (19), selecting a Lyapunov function as follow

$$V = \frac{2k_2}{\alpha+1}|z_1|^{\alpha+1} + \frac{1}{2}z_2^2 + \frac{1}{2}\left(k_1|z_1|^{\frac{\alpha+1}{2}}\operatorname{sgn}(z_1) - z_2\right)^2 \tag{20}$$

We can get that

$$\dot{V} \leq -\frac{\left(\lambda_{\min}\{P\}\right)^{\frac{1-\alpha}{2(\alpha+1)}} \lambda_{\min}\{Q\}}{\lambda_{\min}\{P\}} V^{\frac{3\alpha+1}{2(\alpha+1)}} \tag{21}$$

and

$$0 < \frac{3\alpha+1}{2(\alpha+1)} < 1 \tag{22}$$

where

$$P = \frac{1}{2}\begin{bmatrix} \frac{4k_2}{\alpha+1} + k_1^2 & -k_1 \\ -k_1 & 2 \end{bmatrix}, \quad Q = \frac{k_1}{2}\begin{bmatrix} 2k_2 + k_1^2(\alpha+1) & -k_1(\alpha+1) \\ -k_1(\alpha+1) & (\alpha+1) \end{bmatrix} \tag{23}$$

Therefore, we know that system (19) is finite time stable.

4) Second-order continuous hybrid system:

$$\dot{w}_1 = w_2 - k_1 |w_1|^{\frac{\alpha+1}{2}} \operatorname{sgn}(w_1) - k_3 w_1, \quad \alpha \in (0,1), \quad k_1, k_2 k_3, k_4 > 0 \tag{24}$$
$$\dot{w}_2 = -k_2 |w_1|^{\alpha} \operatorname{sgn}(w_1) - k_4 w_1$$

For second-order continuous nonlinear system (24), selecting a Lyapunov function as follow

$$V = \frac{2k_3}{\alpha+1}|w_1|^{\alpha+1} + k_4 w_1^2 + \frac{1}{2}w_2^2 + \frac{1}{2}\left(k_1|w_1|^{\frac{\alpha+1}{2}}\operatorname{sgn}(w_1) + k_2 w_1 - w_2\right)^2 \tag{25}$$

We can get that

$$\dot{V} \leq -\frac{\lambda_{\min}\{\Omega_1\}}{\sqrt{\lambda_{\max}\{\Pi\}}} V^{0.5} - \frac{\lambda_{\min}\{\Omega_2\}}{\lambda_{\max}\{\Pi\}} V \tag{26}$$

where

$$\Omega_1 = \frac{k_1}{2} \begin{bmatrix} (2k_2+k_1^2(\alpha+1)) & 0 & -k_1(\alpha+1) \\ 0 & (2k_4+k_3^2(\alpha+5)) & -k_3(\alpha+3) \\ -k_1(\alpha+1) & -k_3(\alpha+3) & (\alpha+1) \end{bmatrix},$$

$$\Omega_2 = k_3 \begin{bmatrix} (k_2+k_3^2(\alpha+2)) & 0 & 0 \\ 0 & (k_4+k_3^2) & -k_3 \\ 0 & -k_3 & 1 \end{bmatrix}, \Pi = \begin{bmatrix} \left(\frac{4k_2}{\alpha+1}\right) & k_1k_3 & -k_1 \\ k_1k_3 & (2k_4+k_3^2) & -k_3 \\ -k_1 & -k_3 & 2 \end{bmatrix} \quad (27)$$

Therefore, we know that system (24) is finite time stable.

Parameters: $k_1 = 6, k_2 = 9, k_3 = 10, k_4 = 20, \alpha = 0.2$. The simulation curves of the four systems are shown in Figures 4-6.

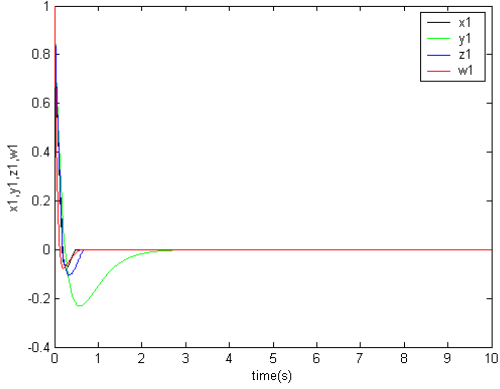

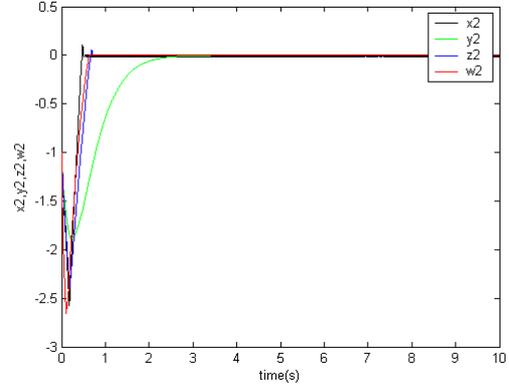

Fig. 4 $x_1$, $y_1$, $z_1$ and $w_1$     Fig. 5 $x_2$, $y_2$, $z_2$ and $w_2$

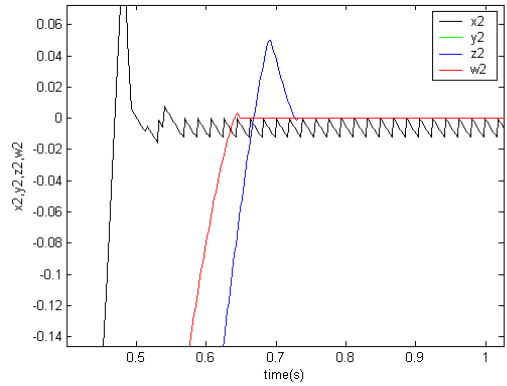

Fig. 6 The magnified figure of Fig. 5

We can find that though $x_1$ is smooth, there exists chattering phenomenon in the output $x_2$ of system (17), because a switching function exists in the second equation of system (17). Therefore, for the robust exact differentiator (7), which has the same structure with system (17), chattering phenomenon in derivative estimation is inevitable. Moreover, no chattering phenomenon exist in the outputs of continuous nonlinear system (19) and continuous hybrid system (24), and the behavior of system (24) is better.

## B. Linear high-gain differentiator [4, 19, 22]

For linear high-gain differentiator

$$\begin{cases} \dot{x}_{21} = x_{22} - \dfrac{a_1}{\tau}(x_{21} - v(t)) \\ \dot{x}_{22} = -\dfrac{a_2}{\tau^2}(x_{21} - v(t)) \end{cases} \quad (28)$$

we know that [4, 19, 22]

$$\lim_{\tau \to 0} x_{21} = 0, \lim_{\tau \to 0} x_{22} = 0 \quad (29)$$

where $a_1, a_2 > 0$.

### B.1 Frequency characteristics of linear high-gain differentiator (28)

In the following, we will analyse the ability of restraining noises for the linear differentiator (28). Laplace transformation of system (28) is

$$\begin{aligned} sX_{21}(s) &= X_{22}(s) - \dfrac{a_1}{\tau}(X_{21}(s) - V(s)) \\ sX_{22}(s) &= -\dfrac{a_2}{\tau^2}(X_{21}(s) - V(s)) \end{aligned} \quad (30)$$

where $X_{21}(s)$, $X_{22}(s)$ and $V(s)$ are respectively Laplace transformations of $x_{21}(t)$, $x_{22}(t)$ and $v(t)$. Therefore, we have the transfer function of signal tracking

$$\frac{X_{21}(s)}{V(s)} = \frac{\dfrac{a_1}{\tau}s + \dfrac{a_2}{\tau^2}}{s^2 + \dfrac{a_1}{\tau}s + \dfrac{a_2}{\tau^2}} \quad (31)$$

and the transfer function of derivative estimation

$$\frac{X_{22}(s)}{V(s)} = \frac{\frac{a_2}{\tau^2}s}{s^2 + \frac{a_1}{\tau}s + \frac{a_2}{\tau^2}} \tag{32}$$

Then, the nature frequency $\omega_n$ is

$$\omega_n = \frac{\sqrt{a_2}}{\tau} \tag{33}$$

and the damping coefficient $\xi$ is satisfied with

$$\xi = \frac{a_1}{2\sqrt{a_2}} \tag{34}$$

For the transfer function (31) and (32), let $s = j\omega$, respectively we have

$$L(\omega) = 20\lg\left|\frac{X_{21}(j\omega)}{V(j\omega)}\right| \approx 0 \quad \text{as} \quad \omega \ll \omega_n \tag{35}$$

$$L(\omega) = 20\lg\left|\frac{X_{21}(j\omega)}{V(j\omega)}\right| \approx -40\lg\frac{\omega}{\omega_n} \quad \text{as} \quad \omega \gg \omega_n \tag{36}$$

and

$$L_d(\omega) = 20\lg\left|\frac{X_{22}(j\omega)}{V(j\omega)}\right| \approx 20\lg\omega \quad \text{as} \quad \omega \ll \omega_n \tag{37}$$

$$L_d(\omega) = 20\lg\left|\frac{X_{22}(j\omega)}{V(j\omega)}\right| \approx 20\lg\frac{\omega_n}{\frac{\omega}{\omega_n}} = 20\lg\omega_n - 20\lg\frac{\omega}{\omega_n} \quad \text{as} \quad \omega \gg \omega_n \tag{38}$$

From the analysis above, linear differentiator can restrain high-frequency noises under the condition that $\tau$ should be selected a suitable value, but not a sufficiently small value. In the following, we will analyse the convergence of linear differentiator (28) by selecting finite parameter $\tau$.

*B. 2 convergence of linear differentiator (28) by finite parameter $\tau$*

Let

$$e_{21} = x_{21} - v(t), e_{22} = x_{22} - \dot{v}(t) \tag{39}$$

then we have the error system as follow:

$$\dot{e} = Ae + B\ddot{v}(t) \tag{40}$$

where

$$A = \begin{bmatrix} -\dfrac{a_1}{\tau} & 1 \\ -\dfrac{a_2}{\tau^2} & 0 \end{bmatrix}, B = \begin{bmatrix} 0 \\ -1 \end{bmatrix} \tag{41}$$

The solution trajectory of the error system is:

$$e(t) = e(0)e^{At} + \int_0^t e^{A(t-\eta)} B d\eta \tag{42}$$

There exist positive constants $\lambda, \sigma_1$, such that

$$\|e(t)\| \le \sigma_1 e^{-\frac{\lambda}{\tau}t} \|e(0)\| + L_2 \int_0^t \|e^{A(t-\eta)}\| d\eta \tag{43}$$

Moreover, the following inequality is satisfied:

$$\|e(t)\| \le \sigma_1 e^{-\frac{\lambda}{\tau}t} \|e(0)\| + \sigma_1 L_2 \int_0^t e^{-\frac{\lambda}{\tau}(t-\eta)} d\eta \tag{44}$$

Therefore, we have

$$\|e(t)\| \le \sigma_1 e^{-\frac{\lambda}{\tau}t} \|e(0)\| + \frac{\tau \sigma_1 L_2}{\lambda} \left(1 - e^{-\frac{\lambda}{\tau}t}\right) \tag{45}$$

Therefore,

$$\lim_{t \to \infty} \|e(t)\| \le \frac{\tau \sigma_1 L_2}{\lambda} = \frac{\sqrt{a_2} \sigma_1 L_2}{\omega_n \lambda} \tag{46}$$

From the analysis above, in order to make $e(t)$ sufficiently small, parameter $\tau$ should be selected sufficiently small. However, suitable nature frequency $\omega_n$ should be selected in order to restrain high-frequency noises. Therefore, parameter $\tau$ cannot be sufficiently small. Thus this linear differentiator exists a static error decided by the selection of parameter $\tau$.

For linear differentiator (28), we also find that when the tracking error is large, i.e., far from the origin, its behavior is strong. On the other hand, when tracking error is small, i.e., around the origin, its behavior is weak.

In general, we want to have the two abilities of linear high-gain differentiator and robust exact differentiator. In the following, we will introduce a global robust exact differentiator, which use a switching function to switch between the two differentiators.

## C. Global robust exact differentiator (GRED)

For the following GRED [28, 29]

$$\begin{cases} \dot{x}_{11} = x_{12} - \lambda_0 |x_{11} - v(t)|^{\frac{1}{2}} \operatorname{sgn}(x_{11} - v(t)) \\ \dot{x}_{12} = -\lambda_1 \operatorname{sgn}(x_{11} - v(t)) \end{cases}, \quad \begin{cases} \dot{x}_{21} = x_{22} - \dfrac{a_1}{\tau}(x_{21} - v(t)) \\ \dot{x}_{22} = -\dfrac{a_2}{\tau^2}(x_{21} - v(t)) \end{cases} \tag{47}$$

The position tracking output $y_1$ and derivative estimation output $y_2$ are respectively

$$y_1 = \alpha_1(e_p) x_{21} + (1 - \alpha_1(e_p)) x_{11} \tag{48}$$

$$y_2 = \alpha_2(e_d) x_{22} + (1 - \alpha_2(e_d)) x_{12} \tag{49}$$

where

$$e_p = x_{11} - x_{21}, \quad e_d = x_{12} - x_{22}$$

$$\alpha_1(e_p) = \begin{cases} 0, & |e_p| < \varepsilon_p - c_p \\ \dfrac{|e_p| - \varepsilon_p + c_p}{c_p}, & \varepsilon_p - c_p \leq |e_p| < \varepsilon_p \\ 1, & |e_p| \geq \varepsilon_p \end{cases}, \quad \alpha_2(e_d) = \begin{cases} 0, & |e_d| < \varepsilon_d - c_d \\ \dfrac{|e_d| - \varepsilon_d + c_d}{c_d}, & \varepsilon_d - c_d \leq |e_d| < \varepsilon_d \\ 1, & |e_d| \geq \varepsilon_d \end{cases} \tag{50}$$

$c_p$ and $c_d$ are the boundary layers respectively used to smoothen the switching functions of signal tracking and derivative estimation, and $\varepsilon_p = K_p \tau, \varepsilon_d = K_d \tau$ with $k_p$ and $k_d$ being the appropriate positive design parameters.

Due to the existence of noises, the outputs of linear differentiator and robust exact differentiator are all not exact. From (9) and (46), we get

$$|e_p| = |x_{11} - x_{21}| = |e_{11} - e_{21}| \leq |e_{11}| + |e_{21}| \leq \mu_0 \varepsilon + \frac{\tau \sigma_1 L_2}{\lambda} \tag{51}$$

$$|e_d| = |x_{12} - x_{22}| = |e_{12} - e_{22}| \leq |e_{12}| + |e_{22}| \leq \mu_1 \varepsilon^{\frac{1}{2}} + \frac{\tau \sigma_1 L_2}{\lambda} \tag{52}$$

It is difficult to select the parameters $k_p$ and $k_d$ when noises exist in signal. It may cause the switching function invalid to switches between the two differentiators. Moreover, though no chattering phenomenon happens for signal tracking, it still exists in derivative estimation because of discontinuous function in the second equation of robust exact differentiator.

In the following, we will present a continuous hybrid differentiator based on a strong Lyapunov function. The differentiator design can not only reduce sufficiently chattering phenomenon of derivative estimation, but also its dynamical performances are improved by adding linear correction terms to the nonlinear ones. Moreover, the robustness will be given by Lyapunov function method.

## 4. Hybrid continuous nonlinear differentiator

### A. Convergence of hybrid continuous nonlinear differentiator

The hybrid continuous differentiator is given as follow:

$$\begin{aligned}\dot{x}_1 &= x_2 - k_1 |x_1 - v(t)|^{\frac{\alpha+1}{2}} \operatorname{sgn}(x_1 - v(t)) - k_2(x_1 - v(t)) \\ \dot{x}_2 &= -k_3 |x_1 - v(t)|^{\alpha} \operatorname{sgn}(x_1 - v(t)) - k_4(x_1 - v(t))\end{aligned} \quad (53)$$

**Theorem 1:** For continuous hybrid differentiator (53) and bounded second-order differentiable signal $v(t)$, there exist constants $k_i > 0$ $(i=1,\cdots,4)$, $t_s > 0$ and $0 < \alpha < 1$ such that

$$\|\varsigma\|_2 \leq \left(L_2 \|\Gamma_1\|_2 / (\lambda_{\min}\{\Omega_1\})\right)^{\frac{\alpha+1}{2\alpha}} \quad (54)$$

for $t \geq t_s$. Where $|\ddot{v}(t)| \leq L_2$, and

$$\varsigma = \left[|e_1|^{\frac{\alpha+1}{2}} \operatorname{sgn}(e_1) \quad e_1 \quad e_2\right]^T, \quad e_1 = x_1 - v(t), \quad e_2 = x_2 - \dot{v}(t), \quad \Gamma_1 = [k_1 \quad k_2 \quad -2] \quad (55)$$

$$\Omega_1 = \frac{k_1}{2} \begin{bmatrix} (2k_3 + k_1^2(\alpha+1)) & 0 & -k_1(\alpha+1) \\ 0 & (2k_4 + k_2^2(\alpha+5)) & -k_2(\alpha+3) \\ -k_1(\alpha+1) & -k_2(\alpha+3) & (\alpha+1) \end{bmatrix} \quad (56)$$

In fact, $L_2 \|\Gamma_1\|_2 / (\lambda_{\min}\{\Omega_1\}) < 1$ can be obtained in differentiator (53) and 0<$\alpha$<1 is selected sufficiently small, ($\alpha$+1)/(2$\alpha$) is sufficiently large, therefore, $\|\varsigma\|_2$ becomes sufficiently in a finite time. This will be given in the following proof.

**Proof:** Let

$$e_1 = x_1 - v(t), e_2 = x_2 - \dot{v}(t) \tag{57}$$

The error system is

$$\dot{e}_1 = e_2 - k_1 |e_1|^{\frac{\alpha+1}{2}} \mathrm{sgn}(e_1) - k_2 e_1 \tag{58}$$
$$\dot{e}_2 = -k_3 |e_1|^{\alpha} \mathrm{sgn}(e_1) - k_4 e_1 - \ddot{v}(t)$$

The Lyapunov function is selected as

$$V = \frac{2k_3}{\alpha+1}|e_1|^{\alpha+1} + k_4 e_1^2 + \frac{1}{2}e_2^2 + \frac{1}{2}\left(k_1 |e_1|^{\frac{\alpha+1}{2}} \mathrm{sgn}(e_1) + k_2 e_1 - e_2\right)^2 \tag{59}$$

Therefore, we have

$$V = \varsigma^{\mathrm{T}} \Pi \varsigma \tag{60}$$

where

$$\varsigma = \begin{bmatrix} |e_1|^{\frac{\alpha+1}{2}} \mathrm{sgn}(e_1) & e_1 & e_2 \end{bmatrix}^{\mathrm{T}}, \quad \Pi = \begin{bmatrix} \left(\frac{4k_3}{\alpha+1}\right) & k_1 k_2 & -k_1 \\ k_1 k_2 & (2k_4 + k_2^2) & -k_2 \\ -k_1 & -k_2 & 2 \end{bmatrix} \tag{61}$$

Moreover, it satisfies

$$\lambda_{\min}\{\Pi\}\|\varsigma\|_2^2 \leq V \leq \lambda_{\max}\{\Pi\}\|\varsigma\|_2^2 \tag{62}$$

where

$$\|\varsigma\|_2^2 = |e_1|^{\alpha+1} + e_1^2 + e_2^2 \tag{63}$$

The time derivative along the trajectories of the hybrid differentiator is

$$\dot{V} = -|e_1|^{\frac{\alpha-1}{2}} \varsigma^{\mathrm{T}} \Omega_1 \varsigma - \varsigma^{\mathrm{T}} \Omega_2 \varsigma + \ddot{v}(t) \Gamma_1 \varsigma \tag{64}$$

where

$$\Omega_1 = \frac{k_1}{2}\begin{bmatrix} (2k_3 + k_1^2(\alpha+1)) & 0 & -k_1(\alpha+1) \\ 0 & (2k_4 + k_2^2(\alpha+5)) & -k_2(\alpha+3) \\ -k_1(\alpha+1) & -k_2(\alpha+3) & (\alpha+1) \end{bmatrix}, \quad \Omega_2 = k_2\begin{bmatrix} (k_3 + k_2^2(\alpha+2)) & 0 & 0 \\ 0 & (k_4 + k_2^2) & -k_2 \\ 0 & -k_2 & 1 \end{bmatrix}, \quad \Gamma_1 = \begin{bmatrix} k_1 & k_2 & -2 \end{bmatrix} \tag{65}$$

Therefore, we have

$$\dot{V} \leq -|e_1|^{\frac{\alpha-1}{2}} \lambda_{\min}\{\Omega_1\}\|\varsigma\|_2^2 - \lambda_{\min}\{\Omega_2\}\|\varsigma\|_2^2 + L_2 \|\Gamma_1\|_2 \|\varsigma\|_2 \tag{66}$$

From (63) and $0<\alpha<1$, we have

$$|e_1|^{\frac{\alpha-1}{2}} \geq \|\varsigma\|_2^{\frac{\alpha-1}{\alpha+1}} \tag{67}$$

Therefore, from (67), inequality (66) can be written as

$$\dot{V} \leq -\lambda_{\min}\{\Omega_1\}\|\varsigma\|_2^{\frac{\alpha-1}{\alpha+1}}\|\varsigma\|_2^2 - \lambda_{\min}\{\Omega_2\}\|\varsigma\|_2^2 + L_2\|\Gamma_1\|_2\|\varsigma\|_2 \tag{68}$$

Therefore, we have

$$\dot{V} \leq -\left(\lambda_{\min}\{\Omega_1\}\|\varsigma\|_2^{\frac{2\alpha}{\alpha+1}} - L_2\|\Gamma_1\|_2\right)\|\varsigma\|_2 - \lambda_{\min}\{\Omega_2\}\|\varsigma\|_2^2 \tag{69}$$

From (62), we get

$$\frac{V}{\lambda_{\max}\{\Pi\}} \leq \|\varsigma\|_2^2 \leq \frac{V}{\lambda_{\min}\{\Pi\}} \tag{70}$$

Therefore, we have

$$\dot{V} \leq -\left(\lambda_{\min}\{\Omega_1\}\|\varsigma\|_2^{\frac{2\alpha}{\alpha+1}} - L_2\|\Gamma_1\|_2\right)\frac{V^{0.5}}{\sqrt{\lambda_{\max}\{\Pi\}}} - \frac{\lambda_{\min}\{\Omega_2\}}{\lambda_{\max}\{\Pi\}}V \tag{71}$$

If $\lambda_{\min}\{\Omega_1\}\|\varsigma\|_2^{\frac{2\alpha}{\alpha+1}} > L_2\|\Gamma_1\|_2$, error system (58) is finite-time stable. Therefore, we have

$$\|\varsigma\|_2 \leq \left(L_2\|\Gamma_1\|_2 / \lambda_{\min}\{\Omega_1\}\right)^{\frac{\alpha+1}{2\alpha}} \tag{72}$$

for $t \geq t_s$. In fact, we can select $k_i, i=1,\cdots,4$ such that $L_2\|\Gamma_1\|_2/\lambda_{\min}\{\Omega_1\} < 1$. Because $0 < \alpha < 1$ is sufficiently small, $(\alpha+1)/(2\alpha)$ is sufficiently large, $\|\varsigma\|_2$ is sufficiently small in a finite time. ∎

For (66), we can find that when the tracking error is large, i.e., far from the origin, $-|e_1|^{\frac{\alpha-1}{2}}\lambda_{\min}\{\Omega_1\}\|\varsigma\|_2^2$ is small, and $-\lambda_{\min}\{\Omega_2\}\|\varsigma\|_2^2$ is large, therefore, the behavior of $-\lambda_{\min}\{\Omega_2\}\|\varsigma\|_2^2$ is strong, and $\lambda_{\min}\{\Omega_2\}$ is decided by the parameters $k_2$ and $k_4$ of linear part. On the other hand, around the origin, $-|e_1|^{\frac{\alpha-1}{2}}\lambda_{\min}\{\Omega_1\}\|\varsigma\|_2^2$ is large, and $-\lambda_{\min}\{\Omega_2\}\|\varsigma\|_2^2$ is small, so the behavior of $-|e_1|^{\frac{\alpha-1}{2}}\lambda_{\min}\{\Omega_1\}\|\varsigma\|_2^2$ is strong. Therefore, the behaviors are strong during the whole convergences.

## B. Robustness analysis for hybrid continuous differentiator

**Theorem 2:** For hybrid differentiator (53), if there exist a noise in signal $v(t)$, i.e., $v(t) = v_0(t) + \delta(t)$, where $v_0(t)$ is the desired second-order derivable signal, and $\delta(t)$ is a bounded noise and is satisfied with $|\delta(t)| \leq \varepsilon$, then, the following inequality is established in finite time

$$\|\varsigma\|_2 \leq \max\left\{\frac{\Psi_1(\varepsilon)}{\lambda_{\min}\{\Omega_1\} - L_2\|\Gamma_1\|_2}, \frac{\Psi_2(\varepsilon)}{\lambda_{\min}\{\Omega_2\}}\right\} \tag{73}$$

where

$$\Psi_1(\varepsilon) = \left[2k_3 + \frac{k_1(\alpha+1)}{2}\|\Gamma_2\|_2\right]\left(k_1 2^{\frac{1-\alpha}{2}} \varepsilon^{\frac{\alpha+1}{2}} + k_2\varepsilon\right)$$

$$\Psi_2(\varepsilon) = \left[(2k_4 + k_2\|\Gamma_2\|_2)\left(k_1 2^{\frac{1-\alpha}{2}} \varepsilon^{\frac{\alpha+1}{2}} + k_2\varepsilon\right) + (1 + \|\Gamma_2\|_2)(k_3 2^{1-\alpha} \varepsilon^\alpha + k_4\varepsilon)\right]$$

$$\Omega_1 = \frac{k_1}{2}\begin{bmatrix}(2k_3 + k_1^2(\alpha+1)) & 0 & -k_1(\alpha+1) \\ 0 & (2k_4 + k_2^2(\alpha+5)) & -k_2(\alpha+3) \\ -k_1(\alpha+1) & -k_2(\alpha+3) & (\alpha+1)\end{bmatrix}, \quad \Omega_2 = k_2\begin{bmatrix}(k_3 + k_2^2(\alpha+2)) & 0 & 0 \\ 0 & (k_4 + k_2^2) & -k_2 \\ 0 & -k_2 & 1\end{bmatrix},$$

$$\Gamma_1 = [k_1 \quad k_2 \quad -2], \quad \Gamma_2 = [k_1 \quad k_2 \quad -1], \quad |\ddot{v}_0(t)| \leq L_2 \tag{74}$$

**Proof:** Let

$$e_1 = x_1 - v_0(t), e_2 = x_2 - \dot{v}_0(t) \tag{75}$$

The error system is

$$\dot{e}_1 = e_2 - k_1|e_1 - \delta|^{\frac{\alpha+1}{2}}\operatorname{sgn}(e_1 - \delta) - k_2 e_1 + k_2 \delta$$
$$\dot{e}_2 = -k_3|e_1 - \delta|^\alpha \operatorname{sgn}(e_1 - \delta) - k_4 e_1 + k_4 \delta - \ddot{v}_0(t) \tag{76}$$

Let

$$\Delta_1 = -|e_1 - \delta|^{\frac{\alpha+1}{2}}\operatorname{sgn}(e_1 - \delta) + |e_1|^{\frac{\alpha+1}{2}}\operatorname{sgn}(e_1)$$
$$\Delta_2 = -|e_1 - \delta|^\alpha \operatorname{sgn}(e_1 - \delta) + |e_1|^\alpha \operatorname{sgn}(e_1) \tag{77}$$

Therefore, from Assumption 1 and Remark 1, we have

$$|\Delta_1| \leq 2^{\frac{1-\alpha}{2}}|\delta|^{\frac{\alpha+1}{2}} \leq 2^{\frac{1-\alpha}{2}} \varepsilon^{\frac{\alpha+1}{2}}$$
$$|\Delta_2| \leq 2^{1-\alpha}|\delta|^\alpha \leq 2^{1-\alpha}\varepsilon^\alpha \tag{78}$$

The Lyapunov function is selected as

$$V = \frac{2k_3}{\alpha+1}|e_1|^{\alpha+1} + k_4 e_1^2 + \frac{1}{2}e_2^2 + \frac{1}{2}\left(k_1|e_1|^{\frac{\alpha+1}{2}}\operatorname{sgn}(e_1) + k_2 e_1 - e_2\right)^2 \tag{79}$$

Therefore, we have

$$V = \varsigma^T \Pi \varsigma \tag{80}$$

where

$$\varsigma = \begin{bmatrix} |e_1|^{\frac{\alpha+1}{2}} \operatorname{sgn}(e_1) & e_1 & e_2 \end{bmatrix}^{\mathrm{T}}, \quad \Pi = \begin{bmatrix} \left(\frac{4k_3}{\alpha+1}\right) & k_1 k_2 & -k_1 \\ k_1 k_2 & (2k_4 + k_2^2) & -k_2 \\ -k_1 & -k_2 & 2 \end{bmatrix} \tag{81}$$

Moreover, it satisfies

$$\lambda_{\min}\{\Pi\} \|\varsigma\|_2^2 \le V \le \lambda_{\max}\{\Pi\} \|\varsigma\|_2^2 \tag{82}$$

where

$$\|\varsigma\|_2^2 = |e_1|^{\alpha+1} + e_1^2 + e_2^2 \tag{83}$$

The time derivative of $V$ along the solution of error system (76) is

$$\begin{aligned}
\dot{V} &= -|e_1|^{\frac{\alpha-1}{2}} \varsigma^{\mathrm{T}} \Omega_1 \varsigma - \varsigma^{\mathrm{T}} \Omega_2 \varsigma + \ddot{v}(t) \Gamma_1 \varsigma \\
&\quad + |e_1|^{\frac{\alpha-1}{2}} \left[ 2k_3 |e_1|^{\frac{\alpha+1}{2}} \operatorname{sgn}(e_1)(k_1 \Delta_1 + k_2 \delta) \right] + 2k_4 e_1 (k_1 \Delta_1 + k_2 \delta) + e_2 (k_3 \Delta_2 + k_4 \delta) \\
&\quad + \begin{bmatrix} k_1 & k_2 & -1 \end{bmatrix} \varsigma \left\{ \frac{k_1(\alpha+1)}{2} |e_1|^{\frac{\alpha-1}{2}} (k_1 \Delta_1 + k_2 \delta) + k_2 (k_1 \Delta_1 + k_2 \delta) - (k_3 \Delta_2 + k_4 \delta) \right\} \\
&\le -|e_1|^{\frac{\alpha-1}{2}} \lambda_{\min}\{\Omega_1\} \|\varsigma\|_2^2 - \lambda_{\min}\{\Omega_2\} \|\varsigma\|_2^2 + L_2 \|\Gamma_1\|_2 \|\varsigma\|_2 \\
&\quad + |e_1|^{\frac{\alpha-1}{2}} \left[ 2k_3 + \frac{k_1(\alpha+1)}{2} \|\Gamma_2\|_2 \right] \left( k_1 2^{\frac{1-\alpha}{2}} \varepsilon^{\frac{\alpha+1}{2}} + k_2 \varepsilon \right) \|\varsigma\|_2 \\
&\quad + \left[ (2k_4 + k_2 \|\Gamma_2\|_2) \left( k_1 2^{\frac{1-\alpha}{2}} \varepsilon^{\frac{\alpha+1}{2}} + k_2 \varepsilon \right) + (1 + \|\Gamma_2\|_2)(k_3 2^{1-\alpha} \varepsilon^\alpha + k_4 \varepsilon) \right] \|\varsigma\|_2
\end{aligned} \tag{84}$$

where

$$\Omega_1 = \frac{k_1}{2} \begin{bmatrix} (2k_3 + k_1^2(\alpha+1)) & 0 & -k_1(\alpha+1) \\ 0 & (2k_4 + k_2^2(\alpha+5)) & -k_2(\alpha+3) \\ -k_1(\alpha+1) & -k_2(\alpha+3) & (\alpha+1) \end{bmatrix}, \quad \Omega_2 = k_2 \begin{bmatrix} (k_3 + k_2^2(\alpha+2)) & 0 & 0 \\ 0 & (k_4 + k_2^2) & -k_2 \\ 0 & -k_2 & 1 \end{bmatrix},$$

$$\Gamma_1 = \begin{bmatrix} k_1 & k_2 & -2 \end{bmatrix}, \quad \Gamma_2 = \begin{bmatrix} k_1 & k_2 & -1 \end{bmatrix} \tag{85}$$

Let

$$\Psi_1(\varepsilon) = \left[ 2k_3 + \frac{k_1(\alpha+1)}{2} \|\Gamma_2\|_2 \right] \left( k_1 2^{\frac{1-\alpha}{2}} \varepsilon^{\frac{\alpha+1}{2}} + k_2 \varepsilon \right) \tag{86}$$

$$\Psi_2(\varepsilon) = \left[ (2k_4 + k_2 \|\Gamma_2\|_2) \left( k_1 2^{\frac{1-\alpha}{2}} \varepsilon^{\frac{\alpha+1}{2}} + k_2 \varepsilon \right) + (1 + \|\Gamma_2\|_2)(k_3 2^{1-\alpha} \varepsilon^\alpha + k_4 \varepsilon) \right] \tag{87}$$

Therefore, the differential inequality (84) can be rewritten as

$$\dot{V} \leq -|e_1|^{\frac{\alpha-1}{2}} \lambda_{\min}\{\Omega_1\} \|\varsigma\|_2^2 - \lambda_{\min}\{\Omega_2\} \|\varsigma\|_2^2 + |e_1|^{\frac{\alpha-1}{2}} \Psi_1(\varepsilon) \|\varsigma\|_2 + L_2 \|\Gamma_1\|_2 \|\varsigma\|_2 + \Psi_2(\varepsilon) \|\varsigma\|_2 \tag{88}$$

Suppose there exist positive constants $c_1$ and $c_2$ such that

$$\Psi_1(\varepsilon) < c_1 \|\varsigma\|_2 \text{ and } \Psi_2(\varepsilon) < c_2 \|\varsigma\|_2 \tag{89}$$

Therefore, we have

$$\dot{V} \leq -|e_1|^{\frac{\alpha-1}{2}} \left[\lambda_{\min}\{\Omega_1\} - c_1\right] \|\varsigma\|_2^2 + L_2 \|\Gamma_1\|_2 \|\varsigma\|_2 - \left[\lambda_{\min}\{\Omega_2\} - c_2\right] \|\varsigma\|_2^2 \tag{90}$$

From (83) and $0<\alpha<1$, we have

$$|e_1|^{\frac{\alpha-1}{2}} \geq \|\varsigma\|_2^{\frac{\alpha-1}{\alpha+1}} \tag{91}$$

Therefore, we get

$$\begin{aligned}\dot{V} &\leq -\|\varsigma\|_2^{\frac{\alpha-1}{\alpha+1}} \left[\lambda_{\min}\{\Omega_1\} - c_1\right] \|\varsigma\|_2^2 + L_2 \|\Gamma_1\|_2 \|\varsigma\|_2 - \left[\lambda_{\min}\{\Omega_2\} - c_2\right] \|\varsigma\|_2^2 \\ &= -\left\{\left[\lambda_{\min}\{\Omega_1\} - c_1\right] \|\varsigma\|_2^{\frac{2\alpha}{\alpha+1}} - L_2 \|\Gamma_1\|_2\right\} \|\varsigma\|_2 - \left[\lambda_{\min}\{\Omega_2\} - c_2\right] \|\varsigma\|_2^2\end{aligned} \tag{92}$$

And from (82), we get

$$\frac{V}{\lambda_{\max}\{\Pi\}} \leq \|\varsigma\|_2^2 \leq \frac{V}{\lambda_{\min}\{\Pi\}} \tag{93}$$

Therefore,

$$\dot{V} \leq -\left\{\left[\lambda_{\min}\{\Omega_1\} - c_1\right] \|\varsigma\|_2^{\frac{2\alpha}{\alpha+1}} - L_2 \|\Gamma_1\|_2\right\} \frac{V^{0.5}}{\sqrt{\lambda_{\max}\{\Pi\}}} - \left[\lambda_{\min}\{\Omega_2\} - c_2\right] \frac{V}{\lambda_{\max}\{\Pi\}} \tag{94}$$

If

$$\|\varsigma\|_2 > \left(\frac{L_2 \|\Gamma_1\|_2}{\lambda_{\min}\{\Omega_1\} - c_1}\right)^{\frac{\alpha+1}{2\alpha}} \text{ and } \lambda_{\min}\{\Omega_2\} > c_2 \tag{95}$$

the differential inequality (94) is finite time convergent, and the error system (76) is finite-time stable.

Because $\alpha$ is sufficiently small, we want $\left(\dfrac{L_2 \|\Gamma_1\|_2}{\lambda_{\min}\{\Omega_1\} - c_1}\right)^{\frac{\alpha+1}{2\alpha}}$ also to be sufficiently small, therefore, it is required that $0 < \dfrac{L_2 \|\Gamma_1\|_2}{\lambda_{\min}\{\Omega_1\} - c_1} < 1$. Then, we have

$$c_1 < \lambda_{\min}\{\Omega_1\} - L_2 \|\Gamma_1\|_2 \text{ and } c_2 < \lambda_{\min}\{\Omega_2\} \tag{96}$$

Therefore, from (89), we know that if

$$\|\varsigma\|_2 > \frac{\Psi_1(\varepsilon)}{\lambda_{\min}\{\Omega_1\} - L_2\|\Gamma_1\|_2} \quad \text{and} \quad \|\varsigma\|_2 > \frac{\Psi_2(\varepsilon)}{\lambda_{\min}\{\Omega_2\}} \tag{97}$$

the differential inequality (87) is finite time convergent. Therefore, we can get

$$\|\varsigma\|_2 \leq \max\left\{\frac{\Psi_1(\varepsilon)}{\lambda_{\min}\{\Omega_1\} - L_2\|\Gamma_1\|_2}, \frac{\Psi_2(\varepsilon)}{\lambda_{\min}\{\Omega_2\}}\right\} \tag{98}$$

This concludes the proof. ∎

**Remark 2:** The hybrid continuous differentiator (53) consists of the linear and nonlinear differentiators given respectively as follow:

$$\begin{aligned}\dot{x}_1 &= x_2 - k_2(x_1 - v(t)) \\ \dot{x}_2 &= -k_4(x_1 - v(t))\end{aligned} \tag{99}$$

and

$$\begin{aligned}\dot{x}_1 &= x_2 - k_1|x_1 - v(t)|^{\frac{\alpha+1}{2}}\mathrm{sgn}(x_1 - v(t)) \\ \dot{x}_2 &= -k_3|x_1 - v(t)|^{\alpha}\mathrm{sgn}(x_1 - v(t))\end{aligned} \tag{100}$$

In the following, we will give a theorem about continuous nonlinear differentiator (100).

**Remark 3:** When $\alpha = 0$, we can get a hybrid discontinuous differentiator as follow:

$$\begin{aligned}\dot{x}_1 &= x_2 - k_1|x_1 - v(t)|^{\frac{1}{2}}\mathrm{sgn}(x_1 - v(t)) - k_2(x_1 - v(t)) \\ \dot{x}_2 &= -k_3\,\mathrm{sgn}(x_1 - v(t)) - k_4(x_1 - v(t))\end{aligned} \tag{101}$$

and can get that

$$x_1 = v(t), x_2 = \dot{v}(t) \tag{102}$$

For $t \geq t_s$. If the Lyapunov function

$$V = \frac{2k_3}{\alpha+1}|e_1|^{\alpha+1} + k_4 e_1^2 + \frac{1}{2}e_2^2 + \frac{1}{2}\left(k_1|e_1|^{\frac{\alpha+1}{2}}\mathrm{sgn}(e_1) + k_2 e_1 - e_2\right)^2 \tag{103}$$

is selected, we have the above conclusion.

For linear differentiator (99), it is required that $k_2$ and $k_4$ are selected sufficiently large. In the differentiator (101), a discontinuous switching function sgn( ) exists in the second equation, therefore, there is chattering phenomenon in the output $x_2$ although $x_1$ is smooth. When high-frequency noises exist in the signal, this chattering can magnify high-frequency noises around the origin.

## 5. Frequency analysis of hybrid differentiator

The linearization of hybrid continuous differentiator (53) by describing function is

$$\dot{x}_1 = x_2 - \left(k_1 A^{\frac{\alpha-1}{2}} \rho_1 + k_2\right)(x_1 - v(t))$$
$$\dot{x}_2 = -\left(k_3 A^{\alpha-1} \rho_2 + k_4\right)(x_1 - v(t))$$
(104)

The nature frequency of system (150) is

$$\omega_n = \sqrt{\frac{k_3 \rho_2}{A^{1-\alpha}} + k_4}$$
(105)

and the damping coefficient is

$$\varsigma = \frac{k_1 A^{\frac{\alpha-1}{2}} \rho_1 + k_2}{2\sqrt{k_3 A^{\alpha-1} \rho_2 + k_4}}$$
(106)

In the hybrid nonlinear differentiator (53), $k_4$ can be selected relatively small, and the normal nature frequency is relatively small too. The nonlinear item $k_3\rho_2/A^{1-\alpha}$ is the effective compensation to the linear part, i.e., the sliding mode items can compensate the delay brought by the linear filter. When the magnitude of tracking error magnitude $A$ is relatively large, because $0<1-\alpha<1$, $A^{1-\alpha}$ decreases, and $\omega_n$ increases. When $A$ is relatively small, $\omega_n$ becomes small too, chattering and small high-frequency noises can be reduced. Better dynamic characteristic is obtained.

## 6. Simulations

In the following simulations, we select the function of $2\sin(t)$ as the desired signal $_0v(t)$.

1) Derivative estimation without noise

A. For Levant differentiator (7), the parameters are $\lambda_2=6$, $\lambda_1=28$. Figures 7 and 8 show respectively simulation results of signal tracking and derivative estimation by robust exact differentiator without noise.

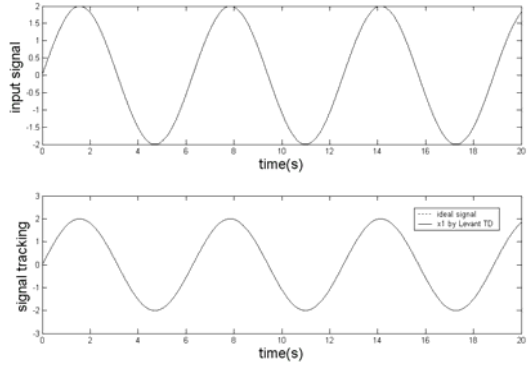
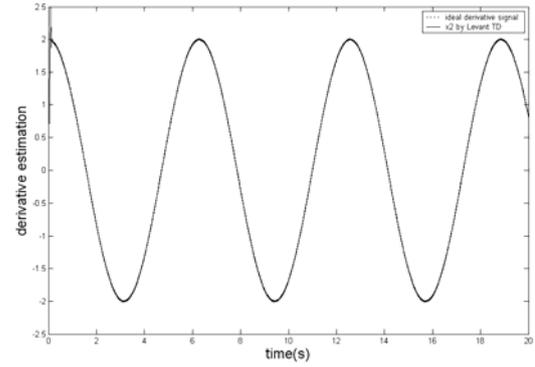

Fig. 7 Signal tracking by robust exact differentiator    Fig. 8-a Derivative estimation by robust exact differentiator

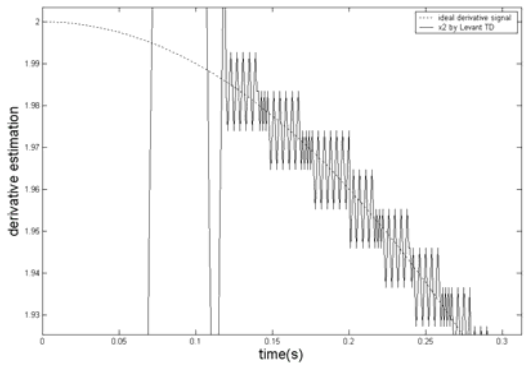
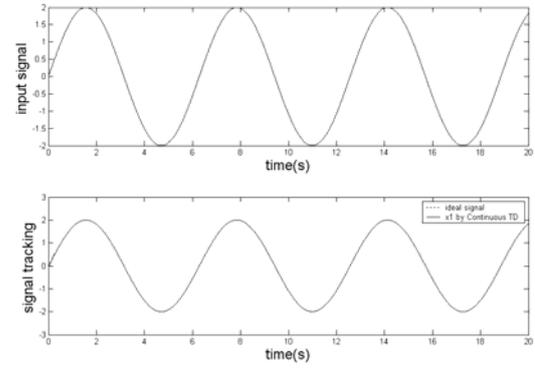

Fig. 8-b The magnified figure of Fig. 8-a    Fig. 9 Signal tracking by hybrid differentiator

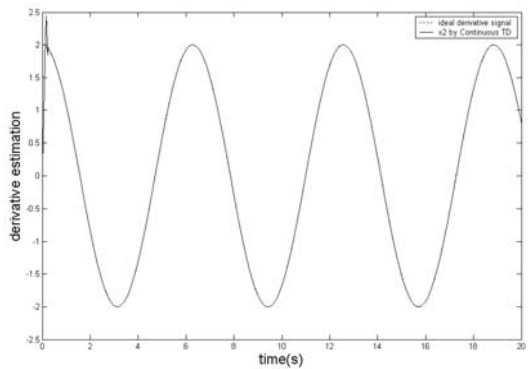
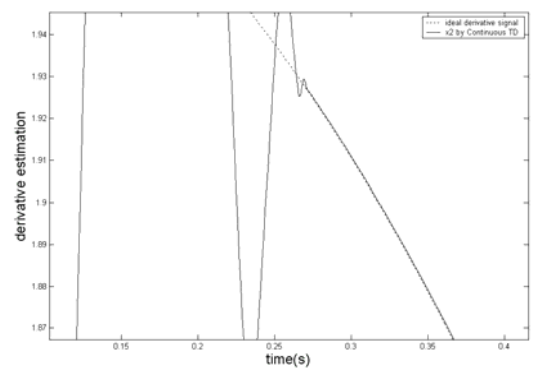

Fig. 10-a Derivative estimation by hybrid differentiator    Fig. 10-b The magnified figure of Fig. 9-a

From the simulations above, we find that the obvious chattering phenomenon happens in robust exact differentiator.

B. For hybrid differentiator, figures 9 and 10 show respectively the simulation results of signal tracking and derivative estimation by hybrid differentiator without noise. From the simulations above, chattering phenomenon is reduced sufficiently and rapid and high-precision tracking can be guaranteed by continuous differentiator.

2) Derivative estimation with noise

A. Robust exact differentiator (7) with $\lambda_2=6$, $\lambda_1=28$. Figures 12 and 13 show the simulation results of signal tracking and derivative estimation respectively by sliding mode differentiator with noises.

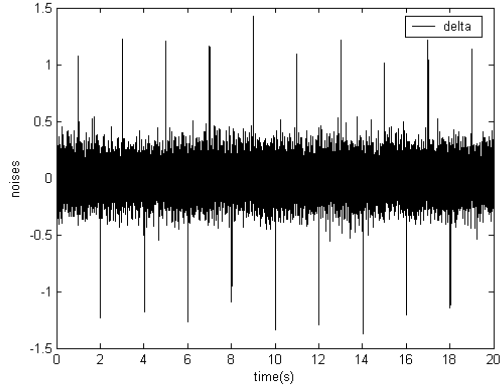

Fig. 11 Noise $\delta(t)$

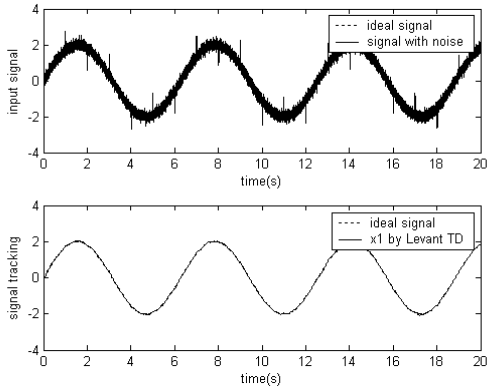
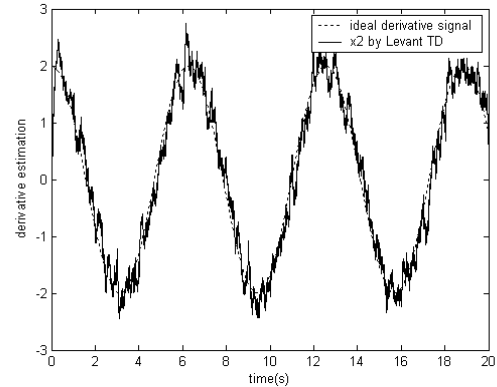

Fig. 12 Signal filtering and tracking by robust exact differentiator    Fig. 13 Derivative estimation by Levant differentiator

B. Global robust exact differentiator (GRED) [28, 29]

$$\begin{cases} \dot{x}_{11} = x_{12} - \lambda_0 |x_{11} - v(t)|^{\frac{1}{2}} \operatorname{sgn}(x_{11} - v(t)) \\ \dot{x}_{12} = -\lambda_1 \operatorname{sgn}(x_{11} - v(t)) \end{cases}, \quad \begin{cases} \dot{x}_{21} = x_{22} - \dfrac{a_1}{\tau}(x_{21} - v(t)) \\ \dot{x}_{22} = -\dfrac{a_2}{\tau^2}(x_{21} - v(t)) \end{cases}$$

The position tracking output $y_1$ and derivative estimation output $y_2$ are respectively

$$y_1 = \alpha_1(e_p) x_{21} + (1 - \alpha_1(e_p)) x_{11}$$

$$y_2 = \alpha_2(e_d) x_{22} + (1 - \alpha_2(e_d)) x_{12}$$

where

$$e_p = x_{11} - x_{21}, \quad e_d = x_{12} - x_{22}$$

$$\alpha_1(e_p) = \begin{cases} 0, & |e_p| < \varepsilon_p - c_p \\ \dfrac{|e_p| - \varepsilon_p + c_p}{c_p}, & \varepsilon_p - c_p \leq |e_p| < \varepsilon_p \\ 1, & |e_p| \geq \varepsilon_p \end{cases}, \quad \alpha_2(e_d) = \begin{cases} 0, & |e_d| < \varepsilon_d - c_d \\ \dfrac{|e_d| - \varepsilon_d + c_d}{c_d}, & \varepsilon_d - c_d \leq |e_d| < \varepsilon_d \\ 1, & |e_d| \geq \varepsilon_d \end{cases}$$

The parameters:

$\lambda_0 = 6, \lambda_1 = 28, \quad \tilde{\lambda}_0 = 0.14, \tilde{\lambda}_1 = 0.2, \tau = 0.1, \quad \varepsilon_p = 10 \times \tau = 1, c_p = 0.5 \times \tau = 0.05, \quad \varepsilon_d = 5 \times \tau = 0.5, c_d = 0.5 \times \tau = 0.05$

Figures 14 and 15 show respectively simulation results of signal tracking and derivative estimation by GRED with noises.

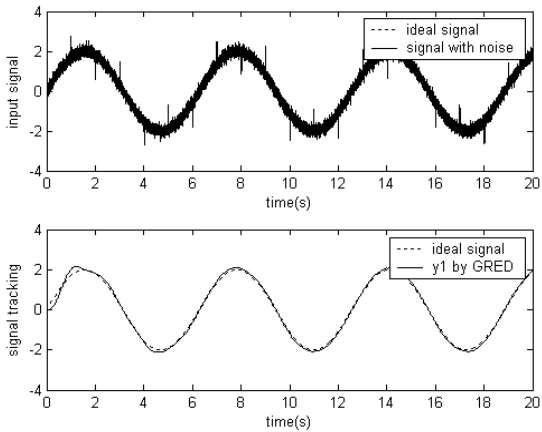
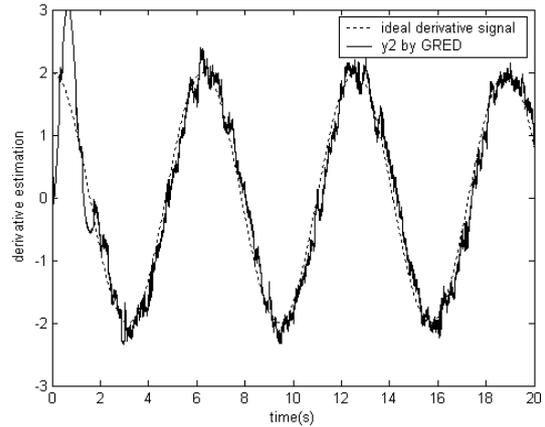

Fig. 14 Signal filtering and tracking by GRED          Fig. 15 Derivative estimation by GRED

D. Hybrid continuous differentiator (53)

In order to restrain peaking phenomenon, we select the following the parameters [30]:

$$\begin{cases} k_2 = 7 \\ k_4 = 25 \end{cases} t \leq 1, \begin{cases} k_2 = 1 \\ k_4 = 8 \end{cases} t > 1$$

$$k_1 = 1, k_3 = 8, \alpha = 0.2$$

We know that when $t \leq 1$, the nature frequency of linear differentiator $\omega_n = 5$, and the damping coefficient $\xi = \dfrac{k_2}{2\sqrt{k_4}} = 0.7$. Figures 16 and 17 show signal tracking and derivative estimation respectively by hybrid continuous differentiator with noises.

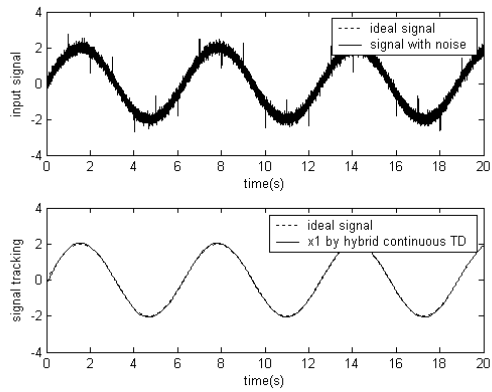 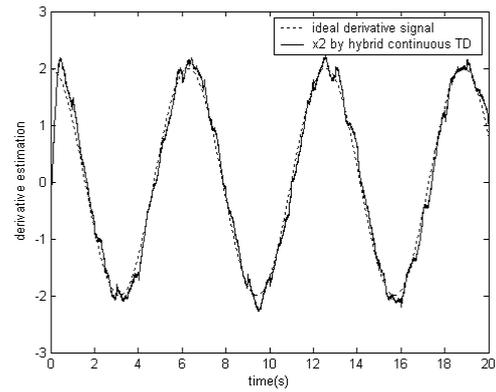

Fig. 16 Signal filtering and tracking by hybrid continuous differentiator

Fig. 17 Derivative estimation by hybrid continuous differentiator

From the simulations above, hybrid continuous differentiator has a better ability of restraining noises and chattering phenomenon. Moreover, small gains can be selected with respect to continuous nonlinear differentiator.

## 7. Conclusion

In this paper, a hybrid continuous differentiator is presented based on a strong Lyapunov function. Because of its continuous structure, and consists of linear and nonlinear parts, not only chattering phenomenon and noises can be reduced sufficiently, but also dynamical performances are improved effectively.